\documentclass[12pt]{article}
\pdfoutput=1
\usepackage{latexsym, graphicx,cite} 
\usepackage{amsmath}
\usepackage{amssymb}

 \newcommand{\arXiv}[1]{\href{http://www.arXiv.org/abs/#1}{#1}}
\usepackage[colorlinks=true, linkcolor=blue, bookmarks=true]{hyperref}

\makeatletter
\renewcommand\section{\@startsection {section}{1}{\z@}%
                                   {-3.5ex \@plus -1ex \@minus -.2ex}
                                   {2.3ex \@plus.2ex}%
                                   {\normalfont\large\bfseries}}
\renewcommand\subsection{\@startsection{subsection}{2}{\z@}%
                                     {-3.25ex\@plus -1ex \@minus -.2ex}%
                                     {1.5ex \@plus .2ex}%
                                     {\normalfont\bfseries}}
\makeatother

\newcommand{\beq}{\begin{equation}}
\newcommand{\eeq}{\end{equation}}
\newcommand{\ber}{\begin{array}}
\newcommand{\eer}{\end{array}}

\newcommand{\del}{\partial}

\newcommand{\de}{\delta}

\newcommand{\eps}{\varepsilon}

\newcommand{\om}{\omega}

\newcommand{\ena}{\end{eqnarray}}
\newcommand{\beqa}{\begin{eqnarray}}
\newcommand{\eeqa}{\end{eqnarray}}
\newcommand{\bea}{\begin{eqnarray}}
\newcommand{\eea}{\end{eqnarray}}

\newcommand{\be}{\begin{equation}}
\newcommand{\ee}{\end{equation}}

\begin{document}
\begin{titlepage}
\begin{flushright}
\phantom{arXiv:yymm.nnnn}
\end{flushright}
\vfill
\begin{center}
{\LARGE\bf Ultraviolet asymptotics for quasiperiodic\vspace{2mm}\\ AdS$_4$ perturbations}   \\
\vskip 15mm
{\large Ben Craps$^a$, Oleg Evnin$^{b,a}$, Puttarak Jai-akson$^{b}$ and Joris Vanhoof$^a$}
\vskip 7mm
{\em $^a$ Theoretische Natuurkunde, Vrije Universiteit Brussel and\\
The International Solvay Institutes\\ Pleinlaan 2, B-1050 Brussels, Belgium}
\vskip 3mm
{\em $^b$ Department of Physics, Faculty of Science, Chulalongkorn University,\\
Thanon Phayathai, Pathumwan, Bangkok 10330, Thailand}

\vskip 3mm
{\small\noindent  {\tt Ben.Craps@vub.ac.be, oleg.evnin@gmail.com,\\ puttarak.jaiakson@gmail.com, Joris.Vanhoof@vub.ac.be}}
\vskip 10mm
\end{center}
\vfill

\begin{center}
{\bf ABSTRACT}\vspace{3mm}
\end{center}

Spherically symmetric perturbations in AdS-scalar field systems of small amplitude $\eps$ approximately periodic on time scales of order $1/\eps^2$ (in the sense that no significant transfer of energy between the AdS normal modes occurs) have played an important role in considerations of AdS stability. They are seen as anchors of stability islands where collapse of small perturbations to black holes does not occur. (This collapse, if it happens, typically develops on time scales of the order $1/\eps^2$.) We construct an analytic treatment of the frequency spectra of such quasiperiodic perturbations, paying  special attention to the large frequency asymptotics. For the case of a self-interacting $\phi^4$ scalar field in a non-dynamical AdS background, we arrive at a fairly complete analytic picture involving quasiperiodic spectra with an exponential suppression modulated by a power law at large mode numbers. For the case of dynamical gravity, the structure of the large frequency asymptotics is more complicated. We give analytic explanations for the general qualitative features of quasiperiodic solutions localized around a single mode, in close parallel to our discussion of the probe scalar field, and find numerical evidence for logarithmic modulations in the gravitational quasiperiodic spectra existing on top of the formulas previously reported in the literature.

\vfill

\end{titlepage}


\section{Introduction}

Numerical investigations in the field of non-linear stability of Anti-de Sitter (AdS) space-times \cite{BR,MRlec} have revealed a patchwork of globally regular and collapsing solutions \cite{DHMS,MRperiod,BLL}. In the corresponding non-linear perturbation theory \cite{FPU,CEV1, CEV2,Buchel:2014xwa,BRM}, one finds a matching competition between pro- and anti-stability features \cite{CEV1,CEV2,Buchel:2014xwa}. In particular, there are powerful sets of selection rules \cite{CEV1,CEV2, Yang, EK} limiting the efficiency of energy transfer between the AdS normal modes in the weakly non-linear regime, both for scalar fields coupled to gravity, as in \cite{BR,MRlec,DHMS,MRperiod,BLL,FPU, CEV1, CEV2,Buchel:2014xwa,BRM}, and probe self-interacting scalar fields in non-dynamical AdS, as in \cite{BKS1,BKS2}.

Ultraviolet dynamics is of crucial importance to the problem of evolution of AdS perturbations. Instability and black hole formation occur through non-linear energy transfer to very short wave-length modes, thus the question of ultraviolet asymptotics in weakly non-linear perturbation theory comes to the fore; see, for instance, the recent paper \cite{BRM}.

An important aspect of the non-linear perturbative AdS dynamics is the existence of periodic and approximately periodic solutions. For example, an algorithm to construct exactly periodic solutions order-by-order in the field amplitude has been presented in \cite{MRperiod} (and extended to massive fields in \cite{Kim:2014ida}). More broadly, one can consider quasiperiodic solutions, in which the AdS normal modes oscillate with constant amplitudes at leading order in the field amplitude on time scales inversely proportional to the square of the field amplitude \cite{FPU}. Since these time scales are precisely the ones on which collapse occurs in numerical experiments, as well as the shortest time scales on which substantial energy drift may occur in the weakly non-linear regime in AdS settings \cite{FPU,CEV1}, the existence of such solutions presents an obstruction to efficient collapse. Islands of stability in the phase space of AdS perturbations are thought to be anchored by such quasiperiodic solutions \cite{Buchel:2014xwa,GMLL}.

At leading order, effectively resummed perturbation theory \cite{FPU,CEV1,CEV2} takes the form of equations describing slow drifts (on time scales inversely proportional to the square of the field amplitude) of the amplitudes and phases of the AdS normal modes. This leading order approximation can be derived by many techniques (multi-scale or renormalization group analysis, averaging) and goes by a few different names in the literature (`two-time framework,' `renormalization flow equations,' `time-averaging,' `resonant approximation'). Quasiperiodic perturbations appear in this context \cite{FPU} as solutions for which the normal mode amplitudes remain constant in time, despite the presence of weak non-linear interactions. This condition can be reduced to an infinite system of cubic algebraic equations, solutions of which were analyzed numerically in \cite{FPU}, revealing exponential damping for high frequency modes. It was pointed out in \cite{CEV1} that the abundance of quasiperiodic solutions is intimately related to selection rules in perturbation theory, which, in turn, give rise to conserved quantities \cite{CEV2}. The solutions were explicitly parametrized by conserved quantities in \cite{Buchel:2014xwa} and numerically shown to be stable in \cite{GMLL}.

Here, we commence our studies of the ultraviolet behavior of the AdS perturbation theory by analyzing the ultraviolet asymptotics of the spectra of such quasiperiodic solutions. We do it both for self-interacting probe scalar field dynamics \cite{BKS2} and for gravitational interactions \cite{FPU}. For the self-interacting probe scalar field, we develop a fairly complete analytical treatment that recovers the essential features of the asymptotic spectra in a way that agrees with direct numerical evaluations. Specifically, one gets amplitude spectra approaching an exponential decay multiplied by the mode number for high-frequency modes. For the gravitational case, both the numerical behavior and the structure of the equations are more involved. We display the asymptotic properties of the coefficients in the equation determining the quasiperiodic spectrum, and explain the general qualitative features of the solutions, such as the existence of two spectral branches previously pointed out numerically \cite{FPU}.  We are unable, however, to draw clear analytic conclusions on the spectral asymptotics from these equations. At the same time, our numerical study of the gravitational quasiperiodic spectra gives evidence for a hierarchy of slow logarithmic modulations superposed on the simple fitting formulas mentioned in \cite{FPU}.

We present our results below by first reviewing (in section 2) the improved perturbation theory in AdS at first non-trivial order, leading to an accurate description of the dynamics on time scales inversely proportional to the square of the field amplitude, as well as the construction of quasiperiodic solutions in this context. We then describe our analysis of the quasiperiodic spectra asymptotics, first for a self-interacting scalar field (section 3), and then for the gravitational case (section 4). Other aspects of ultraviolet asymptotics of AdS dynamics are discussed in our parallel work \cite{CEV3}, where, in particular, we analyze the ultraviolet behavior of the interaction coefficients of the time-averaged equations in arbitrary dimension.

\section{Time-averaged effective theory and quasiperiodic solutions}

We shall briefly review the construction of effective flow equations describing the slow energy transfer between the AdS scalar field modes, starting with the simpler set-up of a self-interacting scalar field \cite{BKS1,BKS2} in AdS$_{d+1}$. A considerably more systematic discussion of these matters (phrased in a context including full gravitational non-linearities) can be found in \cite{CEV1,CEV2}.

We start with the global AdS$_{d+1}$ metric written as
\begin{equation}
ds^{2}=\frac{1}{\cos^2 x}\left(-dt^{2}+dx^{2}+\sin^2 x\,d\Omega_{d-1}^{2}\right),
\end{equation}
where the radial $x$ coordinate runs between 0 and $\pi/2$ and $d\Omega_{d-1}^{2}$ is a $(d-1)$-sphere line element.
The equations of motion for a spherically symmetric probe scalar field $\phi(x,t)$ with a $\phi^4/4!$ non-linearity take the form
\beq
-\del_t^2 \phi +\frac{1}{\tan^{d-1}x}\del_x\left(\tan^{d-1}x\,\del_x\phi\right)=\frac{\phi^3}{3!\cos^2 x}.
\label{NLKG}
\eeq
Expanding this in normal modes $\phi(x,t)=\sum_{n}c_{n}(t)e_{n}(x)$ yields
\beq
\ddot c_n +\om_n^2 c_n =\frac{1}{3!} \sum\limits_{jkl} C_{njkl} c_jc_kc_l,
\eeq
where
\begin{equation}
e_{n}(x)=k_{n}(\cos x)^{d}P_{n}^{\left(\frac{d}{2}-1,\frac{d}{2}\right)}\left(\cos(2x)\right)
\qquad\text{with}\qquad
k_{n}=\frac{2\sqrt{n!(n+d-1)!}}{\Gamma\left(n+\frac{d}{2}\right)},
\label{scmode}
\end{equation}
$P_n$ are Jacobi polynomials, $\om_n=d+2n$ is the peculiar integer (fully resonant) AdS$_{d+1}$ frequency spectrum, and
\begin{equation}
C_{ijkl}=\int_{0}^{\pi/2}e_{i}(x)e_{j}(x)e_{k}(x)e_{l}(x)\frac{\tan^{d-1} x}{\cos^2 x}dx
\end{equation}
are the `interaction coefficients' that will play a key role in our discussions below.

One can now zoom in on the small amplitude regime, assuming amplitudes of order $\epsilon$, and switch to the `interaction picture,' introducing slowly varying complex amplitudes $\alpha_k$ (which would be exactly constant if the non-linearities were neglected):
\beq
c_n=\eps\left(\alpha_ne^{i\om_n t}+\bar \alpha_ne^{-i\om_n t}\right),\qquad \dot c_n=i\eps\om_n\left(\alpha_ne^{i\om_n t}-\bar \alpha_ne^{-i\om_n t}\right), 
\eeq
and hence
\beq
2i\om_n\dot\alpha_n = \frac{\eps^2 e^{-i\om_nt}}{3!} \sum\limits_{jkl}C_{njkl}\left(\alpha_je^{i\om_j t}+\bar \alpha_je^{-i\om_j t}\right)\left(\alpha_ke^{i\om_k t}+\bar \alpha_ke^{-i\om_k t}\right)\left(\alpha_le^{i\om_l t}+\bar \alpha_le^{-i\om_l t}\right).
\label{interact}
\eeq
By the standard lore of averaging \cite{CEV2,BKS1}, when $\eps$ is small, only non-oscillating terms on the right-hand side can contribute significantly. This implies that, out of the sum over $jkl$, only terms satisfying the resonance condition $\om_n=\pm\om_j  \pm\om_k\pm\om_l$ survive (with all the three plus-minus signs {\em a priori} independent). Furthermore, the $C$ coefficients turn out to vanish (see footnote 3 of \cite{CEV2} for a simple proof) for all such resonance conditions other than $\om_n+\om_j=\om_k+\om_l$, i.e., $n+j=k+l$ (and similar relations obtained by interchanging $j$ with $k$ or $l$). One thus arrives at the following effective time-averaged theory accurately approximating the original equations on time scales of order $1/\eps^2$:
\beq
2i\om_n\dot\alpha_n = \frac{\eps^2}2\sum\limits_{jk} C_{njk,n+j-k}\bar\alpha_j\alpha_k\alpha_{n+j-k},
\label{flowscalar}
\eeq
where $3/3!=1/2$ accounts for two other contributions interchanging $j$ with one of the two other summation indices in (\ref{interact}).

One may look, following \cite{FPU,BKS2}, for solutions of the time-averaged equation preserving the normal mode amplitudes $|\alpha_k|$, i.e., solutions in which there is no energy flow between the modes, despite the presence of non-linearities.\footnote{The structure of the energy flow can be elucidated by writing (\ref{flowscalar}) in terms of real amplitudes $A_n$ and phases $B_n$, $\alpha_n=A_ne^{iB_n}$, as we do for the gravitational case in (\ref{Eqn:RGA}-\ref{Eqn:RGB}). In the equation for $A_n$, the right-hand side of (\ref{flowscalar}) is converted to a sum of terms proportional to $A_jA_kA_{n+j-k}\sin(B_n+B_j-B_k-B_{n+j-k})$. Each such terms describes the influence of modes $j$, $k$, $n+j-k$ on the amplitude (energy) of mode $n$ via non-linear interactions. The condition we impose below on the phases, $B_n=B_0+n(B_1-B_0)$, forces these terms to vanish {\em individually}, signifying the absence of resonant energy transfer within any given quartet of modes.} A large class of such solutions is given by imposing $\alpha_{n}=a_{n}e^{-i\eps^2\beta_{n}t}$, where $\beta_{n}=\beta_{0}+n(\beta_{1}-\beta_{0})$, and $a_n$, $\beta_0$, $\beta_1$ are time-independent. This ansatz converts the differential equations (\ref{flowscalar}) to a system of algebraic equations for $a_n, \beta_0, \beta_1$:
\begin{equation}
2\omega_{n}a_{n}(\beta_{0}+n(\beta_{1}-\beta_{0}))=\frac{1}{2}\sum\limits_{jk} C_{njk,n+j-k} a_j a_k a_{n+j-k}.
\label{qpscalar}
\eeq
It is our main objective in this paper to examine what we can say analytically about solutions to the quasiperiodicity conditions of this sort for the  time-averaged system. For the self-interacting scalar field, we shall present a rather complete picture of the ultraviolet properties of solutions with spectra localized around a single mode.

We shall also present some considerations for the case with full gravitational interactions and a free massless scalar field \cite{BR}. The ideology of our treatment is very similar though both technical aspects and phenomenology are more complicated and the amount of progress we can make is considerably more modest. In this case, one considers a spherically symmetric scalar field coupled to a dynamical spherically symmetric metric of the form
\begin{equation}\label{eqn:MetricAnsatz}
ds^{2}=\frac{L^{2}}{\cos^{2}x}\left(\frac{dx^{2}}{A}-Ae^{-2\delta}dt^{2}+\sin^{2}x\,d\Omega_{d-1}^{2}\right),
\end{equation}
with $A$ and $\delta$ being functions of $x$ and $t$. Due to spherical symmetry the metric is determined on each constant time slice by the constraint equations in terms of the matter distribution (i.e., the scalar field profile). One hence essentially obtains an effective theory for $\phi$, albeit with non-linearity much more complicated than in (\ref{NLKG}). Improved perturbative expansions can be constructed for this system \cite{FPU, CEV1, CEV2}, though doing this analytically \cite{CEV1,CEV2} requires considerable effort. See, in particular, \cite{CEV2} for a formulation in the language of time-averaging. Once the effective averaged description of the slow energy transfer has been constructed, one can look for quasiperiodic solutions along the lines we described above for the probe scalar field.

\section{Non-linear probe scalar field in AdS}

\subsection{Quasiperiodic solutions dominated by a single mode}\label{scalarqp}

The conditions (\ref{qpscalar}) defining the spectra of quasiperiodic solutions form an infinite system of cubic algebraic equations, clearly beyond direct reach of any analytic methods.

A particular approach to constructing solutions of (\ref{qpscalar}) has been proposed (in the gravitational context) in \cite{FPU}. Namely one looks for solutions with one mode being much larger than any other mode. (There are obvious exact solutions with only one mode excited, and single-mode-dominated solutions can be seen as perturbations of such obvious single-mode solutions.) In \cite{FPU}, such solutions were constructed numerically.  Here, we shall focus on solutions that are strongly localized in mode number, for which, as it turns out, a fairly complete analytic treatment can be given.

Consider first for simplicity a solution dominated by mode 0, namely $a_0=1$ (there is a scaling symmetry in the equation that always allows to set the amplitude of any given mode to 1), $a_1=\delta\ll 1$. One can then see that there is a consistent ansatz $a_n=q_n\delta^n$, which results in the leading powers on the right-hand-side and left-hand-side matching in all of the equations (\ref{qpscalar}). The procedure of constructing the solutions then goes as follows. In each equation (\ref{qpscalar}) one only keeps the leading powers of $\delta$, matching its coefficient on the two sides of the equation. For equation number 0, this defines $\beta_0$, for equation number 1, this defines $\beta_1$, each subsequent equation determines $q_2$, $q_3$, etc. Importantly, restricting to the leading power of $\delta$ (strong localization limit) in each equation truncates the infinite sums in (\ref{qpscalar}) to a finite number of terms. Specifically,
\begin{equation}
2\omega_{k}q_{k}(\beta_{0}+k(\beta_{1}-\beta_{0}))=\frac{1}{2}\sum\limits_{j=0}^k C_{k0j,k-j} q_j q_{k-j}.
\label{q_scalar}
\eeq
(One of the summation indices in (\ref{qpscalar}) had to be set to 0 to ensure a leading, rather than sub-leading, power of $\delta$.) Equation (\ref{q_scalar}) presents a tremendous simplification over (\ref{qpscalar}) and can be analyzed quite thoroughly. We shall do that after examining the large index properties of the $C$ coefficients.

For solutions dominated by mode number $J>0$, a slighly more complicated, but still completely transparent picture emerges. Namely, one assumes $a_J=1$, $a_{J+1}=\delta\ll 1$. The self-consistent ansatz to substitute into (\ref{qpscalar}) is then taken to be $a_k=q_k \delta^{|k-J|}$. With this ansatz, equation number $J$ determines $\beta_0+J(\beta_1-\beta_0)$. Equations number $J+1$ and $J-1$ thereafter determine $\beta_0$, $\beta_1$ and $q_{J-1}$. These equation are effectively quadratic and have two solutions, which explains the two branches seen in \cite{FPU,BKS2}. Proceeding further iteratively, each pair of equations number $J+n$ and $J-n$ with $n\le J$ forms a $2\times 2$ linear system determining $q_{J+n}$ and $q_{J-n}$. Finally, each equation with number greater than $2J$ determines the $q$ with the corresponding number, much like what (\ref{q_scalar}) does for the $J=0$ case. We shall see below how this procedure works explicitly for $J=1$.

We briefly comment on the general nature of the small $\delta$ expansion, of which we are mainly interested in the leading terms. (Systematic investigations of this issue are outside the scope of our present study.) The pattern we have displayed above suggests an asymptotic expansion around $\delta=0$ of the form
\beq
a_n=\de^{|n-J|}\left(q_n+q^{(1)}_n\de+q^{(2)}_n\de^2+\cdots\right).
\label{deexp}
\eeq
Indeed, after the leading order contribution has been determined in the way we have described, subleading terms can be fixed order-by-order by substituting (\ref{deexp}) into (\ref{qpscalar}) and solving the resulting {\em linear} equations.  Importantly, the prefactor $\de^{|n-J|}$ completely drops out when substituted in the equations (the leading order is $\de$-independent, as seen, for example, in (\ref{q_scalar}) for $J=0$). For that reason, the fact that the asymptotic expansions for different $a_n$ start with different powers of $\de$ by no means upsets the consistency of determining subleading corrections simultaneously for all $a_n$. Note that, for corrections of each given order in $\de$, the infinite sums in (\ref{qpscalar}) will truncate to a finite number of terms. Indeed, in (\ref{qpscalar}) written for mode number $n$, the right-hand side contains an infinite sum over $j$ and $k$, with terms bearing factors of $\de^{|j-J|+|k-J|+|n+j-k-J|}$. Since all terms in the power of $\de$ are positive and the first two terms grow as $j$ or $k$ are increased for $j,k>J$, it is evident that, for each given $n$ and $J$, only a finite number of choices of $j$ and $k$ exist that keep the power of $\de$ below any given value. This leaves a finite number of terms out of the sum in (\ref{qpscalar}) at each $n$ and at each order in $\de$.

Asymptotic expansions of the form (\ref{deexp}) strongly suggest that the leading order results we deal with here become as accurate as one needs for sufficiently small $\de$. The corrections in (\ref{deexp}) are uniquely determined by solving linear equations and simply slightly shift the existing leading order spectral branches we have described above.

\subsection{Asymptotics of the interaction coefficients}\label{Canalysis}

We shall now analyze the asymptotic behavior of the $C$ coefficients appearing in (\ref{q_scalar}) for AdS$_4$. The mode functions for global AdS$_{d+1}$ are given by (\ref{scmode}). For large mode numbers, we can use the approximation (see, e.g., \cite{DLMF}): 
\begin{align}
&(\sin x)^{\alpha+\frac{1}{2}}(\cos x)^{\beta+\frac{1}{2}}P_{n}^{(\alpha,\beta)}(\cos(2x)) \nonumber \\
&\simeq2^{2n+\alpha+\beta+1}\frac{\Gamma(n+\alpha+1)\Gamma(n+\beta+1)}{\pi\Gamma(2n+\alpha+\beta+2)}\left\{\cos\left((2n+\alpha+\beta+1)x-\frac{\pi}{4}(2\alpha+1)\right)+\mathcal{O}\left(\frac{1}{n}\right)\right\}
\end{align}
to write the mode functions as
\begin{align}
e_{n}(x)&=\left(\frac{2\sqrt{\Gamma(n+1)\Gamma(n+d)}}{\Gamma\left(n+\frac{d}{2}\right)}\right)(\cos x)^{d}P_{n}^{\left(\frac{d}{2}-1,\frac{d}{2}\right)}\left(\cos(2x)\right) \nonumber \\
&\approx\frac{2\sqrt{\Gamma(n+1)\Gamma(n+d)}}{\sqrt{\pi}\,\Gamma\left(n+\frac{d}{2}+\frac{1}{2}\right)}(\cot x)^{\frac{d-1}{2}}\left\{\cos\left((d+2n)x-\frac{\pi}{4}(d-1)\right)+\mathcal{O}\left(\frac{1}{n}\right)\right\}.
\end{align}
For generic dimension $d$, the approximate expression for the mode functions will diverge near $x=0$, and we will not be able to employ the straightforward approach for extracting the asymptotics of $C$ coefficients that we utilize below for $d=3$. At $d=3$,
\begin{equation}
e_{n}(x)\approx\frac{2}{\sqrt{\pi}}\left(\frac{\cos x}{\sin x}\right)\left\{\sin(\omega_{n}x)+\mathcal{O}\left(\frac{1}{n}\right)\right\}
\end{equation}
does not diverge. 

We pause to note that this formula is closely related to, and could also be easily derived from, the exact simplified expressions for the AdS$_4$ mode functions that appeared in \cite{GMLL} while we were preparing the current article for publication. Those expressions were used in \cite{GMLL} to obtain closed-form (albeit very long) expressions for the interaction coeffients in the gravitational case, and this could also be done for the case of the probe scalar. While the availability of such closed-form expressions for the interaction coefficients in this particular dimension is certainly a welcome development, the simpler analysis of the asymptotics provided here will be sufficient for our purposes in the present paper. In passing, we also note that an efficient recursive way to compute interaction coefficients in arbitrary dimension is provided in \cite{CEV3}.

We shall start by displaying the asymptotic formula for $C$ with four large indices, which is an instructive aside:
\begin{align}
&C_{nmkl}\simeq\frac{16}{\pi^{2}}\int_{0}^{\pi/2}\frac{\sin(\omega_{n}x)\sin(\omega_{m}x)\sin(\omega_{k}x)\sin(\omega_{l}x)}{(\sin x)^{2}}\text{d}x \nonumber \\
&=\frac{1}{\pi}\Big(|-\omega_{n}+\omega_{m}+\omega_{k}+\omega_{l}|+|\omega_{n}-\omega_{m}+\omega_{k}+\omega_{l}|+|\omega_{n}+\omega_{m}-\omega_{k}+\omega_{l}| \nonumber \\
&\quad+|\omega_{n}+\omega_{m}+\omega_{k}-\omega_{l}|-|\omega_{n}+\omega_{m}+\omega_{k}+\omega_{l}|-|\omega_{n}+\omega_{m}-\omega_{k}-\omega_{l}| \nonumber \\
&\quad-|\omega_{n}-\omega_{m}+\omega_{k}-\omega_{l}|-|\omega_{n}-\omega_{m}-\omega_{k}+\omega_{l}|\Big).
\label{Eqn:AsymptoticCoefficients}
\end{align}
In a general number of dimensions, we expect the first powers of the combinations of frequencies in this expression to generalize to power $d-2$, in analogy to the considerations of \cite{CEV3}.

The actual coefficients appearing in (\ref{q_scalar}) have at least one small index, while the other indices may become large asymptotically. For three large indices, we have\footnote{Here and below, we are using the explicit formulas $P^{(a,b)}_{0}(x)=1$ and $P^{(a,b)}_{1}(x)=\frac{1}{2}(2(a+1)+(a+b+2)(x-1))$.}
\begin{equation}
e_{0}(x)=\frac{4\sqrt{2}}{\sqrt{\pi}}(\cos x)^{3}
\end{equation}
and
\begin{align}
C_{nmk0}\simeq&\frac{32\sqrt{2}}{\pi^{2}}\int_{0}^{\pi/2}\frac{\sin(\omega_{n}x)\sin(\omega_{m}x)\sin(\omega_{k}x)(\cos x)^{2}}{\sin x}\text{d}x \nonumber \\
=&\frac{32\sqrt{2}}{\pi^{2}}\times
\begin{cases}
0 & \text{ if }\,\, \text{max}\{n,k,l\}>(n+k+l+2)/2 \\
\pi/16 & \text{ if }\,\, \text{max}\{n,k,l\}=(n+k+l+2)/2 \\
3\pi/16 & \text{ if }\,\, \text{max}\{n,k,l\}=(n+k+l+1)/2 \\
\pi/4 & \text{ otherwise }
\end{cases}
\end{align}
In particular,
\beq
C_{0kj,k-j}\rightarrow\frac{8\sqrt{2}}{\pi}\approx 3.6013.
\eeq 
We proceed with two small and two large indices, using
\begin{equation}
e_{1}(x)=\frac{4\sqrt{6}}{3\sqrt{\pi}}(\cos x)^{3}(3-8(\sin x)^{2}).
\end{equation}
\begin{align}
C_{nm01}\simeq&\frac{128\sqrt{3}}{3\pi^{2}}\int_{0}^{\pi/2}\sin(\omega_{n}x)\sin(\omega_{m}x)(\cos x)^{4}(3-8(\sin x)^{2})\text{d}x \nonumber \\
=&\frac{128\sqrt{3}}{3\pi^{2}}\times
\begin{cases}
\pi/32 & \text{ if }\,\, |m-n|=3 \\
7\pi/64 & \text{ if }\,\, |m-n|=2 \\
5\pi/32 & \text{ if }\,\, |m-n|=1 \,\,\text{ or }\,\, |m-n|=0 \\
0 & \text{ otherwise }
\end{cases}
\end{align}
So in particular $C_{0,1,n,n+1}\rightarrow\frac{20\sqrt{3}}{3\pi}\approx3.6755$.
\begin{align}
C_{nm00}\simeq&\frac{128}{\pi^{2}}\int_{0}^{\pi/2}\sin(\omega_{n}x)\sin(\omega_{m}x)(\cos x)^{4}\text{d}x \nonumber \\
=&\frac{128}{\pi^{2}}\times
\begin{cases}
\pi/64 & \text{ if }\,\, |m-n|=2 \\
\pi/16 & \text{ if }\,\, |m-n|=1 \\
3\pi/32 & \text{ if }\,\, |m-n|=0 \\
0 & \text{ otherwise }
\end{cases}
\end{align}
So in particular $C_{0,0,n,n}\rightarrow\frac{12}{\pi}\approx3.8197$.

These observations for large mode number limits taken in different directions in the mode number space reveal a very simple plateau-like topography, see Fig.~\ref{Cplot}. Almost everywhere in the large index value region, with one index kept equal to 0 according to (\ref{q_scalar}), the plot of $C$ is completely flat at a fixed height of $8\sqrt{2}/{\pi}$. If one goes to the border regions of this plateau, where one of the three large indices becomes small, the plot rises somewhat, but neither is the rise numerically significant (only a few percent), nor is it statistically significant when appearing inside the sums, as in (\ref{q_scalar}), since this rise is only attained for special, non-generic values of the indices. In the end, the large mode-number asymptotics of the solutions to (\ref{q_scalar}) can be characterized almost entirely in terms of the single `plateau' value of $C$, i.e., $8\sqrt{2}/{\pi}$. We shall see below how this works explicitly.
\begin{figure}[t]
\centering
\includegraphics[width=12cm]{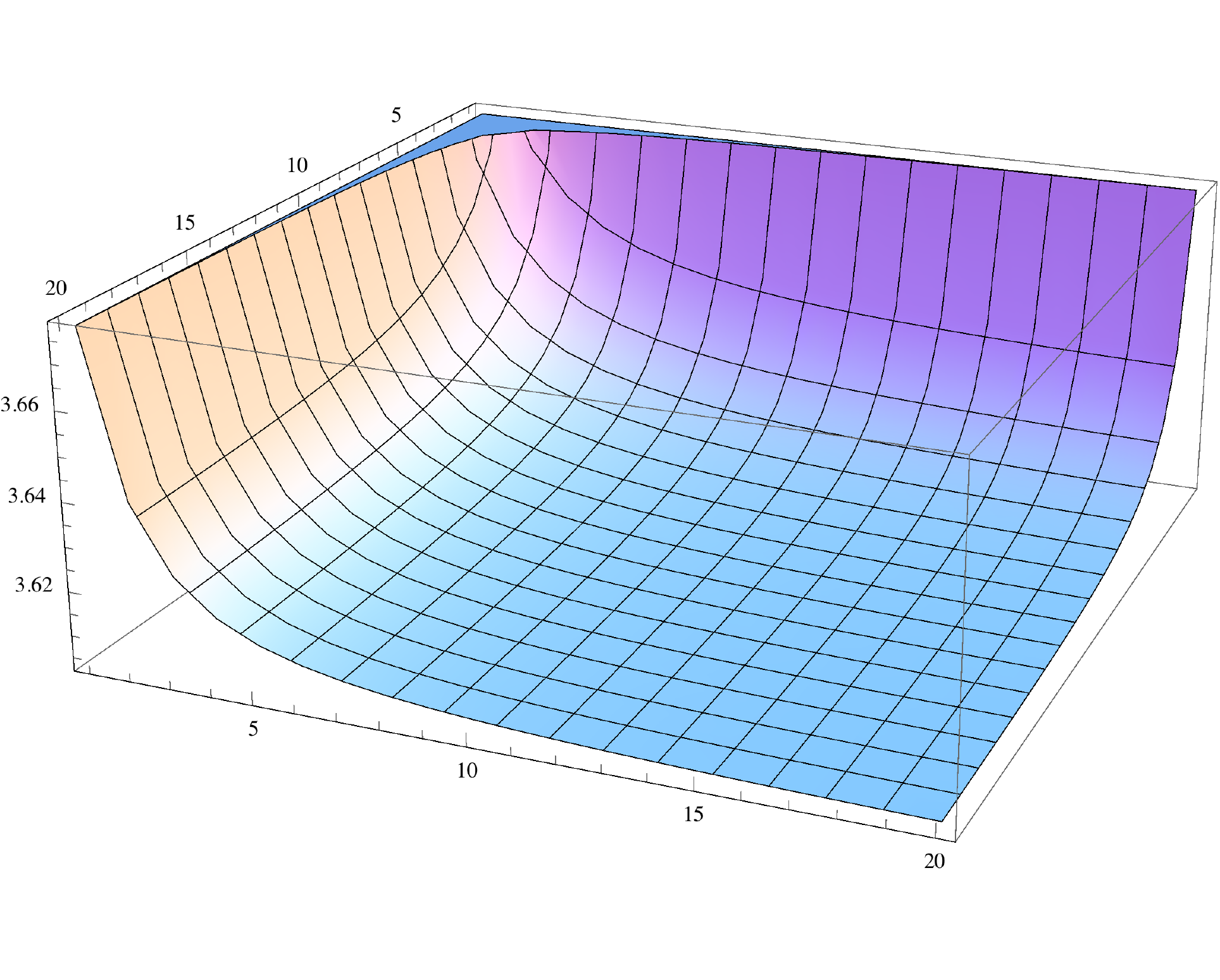}\vspace{-1.2cm}
\caption{$C_{0,n+m,n,m}$ plotted as a function of $n$ and $m$.}
\label{Cplot}
\end{figure}

\subsection{Solutions dominated by mode 0: asymptotic estimates}

Armed with asymptotic expressions for the $C$ coefficients, we can start our analysis of the large mode-number limit of the solutions to (\ref{q_scalar}). As we already explained, one assumes in (\ref{qpscalar}) $a_{0}=1$, $a_{1}=\delta\ll 1$ and $a_{n}=q_{n}\delta^{n}$. From the equations number 0 and 1 in (\ref{qpscalar}), one obtains
\begin{equation}
\beta_{0}=\frac{C_{0000}}{4d},
\qquad
\beta_{1}=\frac{C_{0011}}{2(d+2)}.
\end{equation}
For $d=3$, this becomes
\beq
\beta_{0}=\frac{7}{6\pi},\qquad \beta_{1}=\frac{19}{15\pi}.
\eeq
At higher orders in $\de$, one has to solve (\ref{q_scalar}). Note that the $C$ coefficients inside the sum in (\ref{q_scalar}) are nearly constant at large $k$, except for the endpoints of the sum, where they vary slightly. Since there are considerably more terms in the interior of the sum than near the endpoints, it is natural to assume that the equation can be approximated well by simply taking $C=8\sqrt{2}/\pi$ (the plateau value we have identified above), solving the equation and then checking that our assumption has been justified. It is likewise natural to assume that the term proportional to $k^2$ in the coefficient of $q_k$ dominates the lower powers of $k$. All these assumptions can, of course, be explicitly checked by an {\em a posteriori} comparison to numerical evaluations of the solutions to (\ref{q_scalar}).

Applying the above simplifications to (\ref{q_scalar}), one obtains:
\beq
4k^2(\beta_{1}-\beta_{0})q_{k}=\frac{4\sqrt{2}}{\pi}\,\,\sum\limits_{j=1}^{k-1} q_j q_{k-j},\qquad\mbox{or}\qquad \frac{k^2}{10}q_{k}=\sqrt{2}\,\sum\limits_{j=1}^{k-1} q_j q_{k-j}.
\label{largekmode0}
\eeq
One can assume the simple asymptotic behavior $q_{n}=Qn^{\gamma}e^{\kappa n}$, and check for what values of the parameters $\kappa$, $\gamma$, $Q$ it is compatible with (\ref{largekmode0}).  We find that $\kappa$ completely drops out from (\ref{largekmode0}). It therefore cannot be determined by simple asymptotic matching and requires solving (\ref{q_scalar}) iteratively mode-by-mode. After $\kappa$ has dropped out, one is left with
\begin{equation}
k^{\gamma+2}=10\sqrt{2}\,Q\sum_{j=1}^{k-1}j^{\gamma}(k-j)^{\gamma}.
\label{Qandgamma}
\end{equation}
For large $k$, we can approximate the sum by an integral over $\zeta=j/k$:
\begin{equation}
\sum_{j=1}^{k-1}j^{\gamma}(k-j)^{\gamma}\simeq k^{2\gamma+1}\int_{0}^{1}\text{d}\zeta\,\zeta^{\gamma}(1-\zeta)^{\gamma}.
\end{equation}
Matching the leading powers on both sides of the equation fixes $\gamma=1$. Thereafter, (\ref{Qandgamma}) fixes $Q=3\sqrt{2}/10$.
We thus expect the asymptotic spectrum to behave as
\begin{equation}
a_{n}\simeq\frac{3\sqrt{2}}{10}\,n\left(\delta e^{\kappa}\right)^{n}.
\label{UVscalar0}
\end{equation}
Note that $e^{\kappa}$ cannot be absorbed in $\delta$ since the normalization of $\delta$ is fixed at the infrared end of the spectrum by imposing $a_1=\delta$. We have checked numerically that (\ref{UVscalar0}) accurately approximates exact solutions to (\ref{q_scalar}). In particular, we have verified that no logarithmic modulations are present in addition to the power-law modulation in (\ref{UVscalar0}); we will explain the method we used in section \ref{sec:mode1}. The situation will not be similarly simple when we turn to the gravitationally interacting case.

\subsection{Solutions dominated by mode 0: systematic analysis}

We have explained above that natural simplifying assumptions (which can be verified {\em a posteriori} by direct comparison with numerics, even in the absence of rigorous error bounds) reduce the quasiperiodicity conditions (\ref{q_scalar}) for solutions dominated by mode 0 to (\ref{largekmode0}). We have then verified that the asymptotics (\ref{UVscalar0}) is analytically consistent with the simplified equation (\ref{largekmode0}), and also provides a valid approximation to the full numerical solutions of (\ref{q_scalar}).

At a practical level, one could rest content with (\ref{UVscalar0}). However, it would be better to have a more explicit analytic control over how this asymptotics emerges. (Note that being asymptotically consistent with an equation does not, strictly speaking, imply that solutions to this equation, with initial data specified in the infrared, approach asymptotics of this form.) Equation (\ref{largekmode0}) provides a relatively simple iterative definition of $q_n$ via a quadratic recursive relation reminiscent of the one for the Catalan numbers. It can be analyzed quite exhaustively, with a curious mathematical structure emerging, which is what we shall explain next.

To gain further analytic insights in (\ref{largekmode0}), we introduce a generating function
\beq
q(\lambda)=1+e^{-\lambda}+\sum_{n=2}^\infty q_n e^{-\lambda n}.
\label{qdef}
\eeq
(\ref{largekmode0}), together with the initial conditions $q_0=q_1=1$, is then equivalent to
\beq
q''=e^{-\lambda}+10\sqrt{2}(q-1)^2.
\label{runaway}
\eeq
This equation has an obvious mechanical interpretation: a particle is moving in the $q$-space and $\lambda$-time with an external force $e^{-\lambda}$ and a potential $-10\sqrt{2}(q-1)^3/3$. One is looking for a special solution of this equation that stops at $\lambda=+\infty$ at the inflection point of the potential, i.e., $q=1$. One-dimensional mechanics in a cubic potential is well-known to display a run-away behavior in finite time (all solutions run down the potential slope to $q=+\infty$ at a certain finite moment of time). In the vicinity of that run-away moment, which we shall call $\kappa$, the $e^{-\lambda}$ term in (\ref{runaway}) is negligible and one can integrate the equation to obtain
\beq
q(\lambda)\approx \frac{3\sqrt{2}}{10\,(\lambda-\kappa)^2}.
\label{runsoln}
\eeq
Of course, the particular value of the run-away moment $\kappa$ is sensitive to the entire evolution and requires solving (\ref{runaway}) exactly, which cannot be done analytically.

What have we learned? $q$ defined by (\ref{qdef}) is an analytic function of $z=e^{-\lambda}$ with a double pole singularity at $z=e^{-\kappa}$. $q_n$ are the coefficients of the Taylor series expansion of this function around $z=0$. It is generally understood that high order behavior of the Taylor series is governed by the singularities of the expanded function in the complex plane. Darboux' theorem (see, e.g., \cite{darboux}) gives a precise statement of this correspondence. In particular, a double pole at $z=e^{-\kappa}$ precisely corresponds to $q_n\sim n e^{\kappa n}$. Furthermore, the coefficient of (\ref{runsoln}) matches the coefficient of (\ref{UVscalar0}). Overall, our systematic analysis of (\ref{largekmode0}) based on generating functions completely corroborates the simple asymptotic estimates of the previous section.

\subsection{Solutions dominated by mode 1}\label{sec:mode1}

As a proof-of-concept demonstration, we shall now describe how to analyze solutions dominated by mode 1. Compared to the solutions dominated by mode 0 considered above, this case presents the feature of having two spectral branches, originally seen for gravity in the numerical simulations of \cite{FPU}. Extension to solutions dominated by still higher modes, on the other hand, is a pure technicality.

We assume a spectrum $a_1 = 1$, $a_2=\delta\ll1$ and $a_n= q_n\delta^{|n-1|}$, substitute it in (\ref{qpscalar}) and retain only the leading powers of $\delta$ in each equation.  (\ref{qpscalar}) with $n=0,1,2$ gives
\begin{align}
&2\omega_0q_0\beta_0 = \frac{1}{2}\left(2C_{1001}q_0+C_{1102}\right),\\
&2\omega_1\beta_1 =  \frac{1}{2}C_{1111},\\
&2\omega_2(2\beta_1-\beta_0)  = \frac{1}{2}(C_{1120}q_0+2C_{1221}).
\end{align}
For $d=3$, $\beta_1 = {19}/{18\pi}$. Solving the two remaining equations simultaneously leads to
\begin{equation}
q_0^2 + \frac{185\sqrt{6}}{132}q_0 + \frac{7}{3} = 0,
\end{equation} 
with two solutions, corresponding to two spectral branches of the type seen numerically in \cite{FPU}:
\begin{align}
&q_0 = \frac{-185\sqrt{6}+\sqrt{42726}}{264},  \ \ \ \beta_0 = \frac{347-\sqrt{7121}}{252\pi} \label{branch1} ,\\
&q_0 = \frac{-185\sqrt{6}-\sqrt{42726}}{264},  \ \ \ \beta_0 = \frac{347+\sqrt{7121}}{252\pi}.\label{branch2}
\end{align}
The remaining equations reduce to
\begin{equation} \label{eq:6}
4\omega_k(\beta_0+k(\beta_1-\beta_0))q_k = q_0\displaystyle{\sum_{j=1}^{k-1}}C_{0,k,j,k-j}q_jq_{k-j}+\displaystyle{\sum_{j=1}^{k}}C_{1,k,j,k+1-j}q_jq_{k+1-j}.
\end{equation} 

In the limit of large $k$, we found that, for $d=3$, the interaction coefficients are approximately constant: 
\beq
C_{0,k,j,k-j} \approx \frac{8\sqrt{2}}{\pi} \equiv C_0 ,\qquad\mbox{and}\qquad C_{1,k,j,k+1-j} \approx \frac{20}{\pi} \equiv C_1.
\eeq
By evoking similar logic to what we have already used in our mode 0 derivations (without recapitulating all the subtleties), we assume an asymptotic ansatz $q_n = Qn^\gamma e^{\kappa n }$ and substitute it in (\ref{eq:6}), obtaining
\begin{equation} \label{eq:8}
{4\omega_k}(\beta_0+k(\beta_1-\beta_0))k^\gamma = q_0C_0Q\displaystyle{\sum_{j=1}^{k-1}j^\gamma(k-j)^\gamma}+C_1Qe^{\kappa}\displaystyle{\sum_{j=1}^{k}j^\gamma(k+1-j)^\gamma}.
\end{equation}
Matching the leading power of $k$ on both sides of this equation results in $\gamma = 1$. We also obtain the following relation between $Q$ and $\kappa$,
\begin{equation}
Q = \frac{48(\beta_1-\beta_0)}{q_0C_0+C_1e^\kappa},
\label{Qkappa}
\end{equation}
with the specific numerical values for $\beta_0$, $\beta_1$, $q_0$, $C_0$, $C_1$ given above.

We then solve (\ref{eq:6}) numerically and fit the results to $q_n=Qn^\gamma e^{\kappa n }(1+\frac{c}{n}+\frac{c_2}{n^2})$. This is done by fitting 5 adjacent points to the above expression, and then moving these 5 points to higher and higher values of $n$ in hope to see convergence. Quick convergence does, in fact, follow. For the solution (\ref{branch1}), we get $\gamma=1$, $\kappa\approx-0.06$, $Q\approx 0.0846$. For the solution (\ref{branch2}), we get $\gamma=1$, $\kappa\approx -0.648$, $Q\approx 1.75$. Both sets of values of $\kappa$ and $Q$
are in agreement with (\ref{Qkappa}). We therefore conclude that our analytic treatment of the large mode-number limit provides a sound perspective. Attempts to augment the asymptotics with logarithmic modulations, for example, a factor $(\ln n)^{\gamma_1}$, and obtaining estimates for $\gamma_1$, result in a quick convergence of $\gamma_1$ to 0 as the mode number increases. This indicates the absence of logarithmic modulations, in accordance with our analytic results. We shall now turn to the case of gravitational interactions, where logarithmic modulations are, in fact, present.

\section{Gravitational interactions}

\subsection{Gravitational quasiperiodic solutions}

In the introduction, we have briefly explained that, compared to the self-interactive probe scalar field we have considered above, the effective time-averaged description for the full gravitational non-linearities is conceptually quite similar, but technically much more complicated. The averaged equations, as derived \cite{CEV2}, are of the form (we are now using real amplitudes $A_n$ and phases $B_n$, related to the complex amplitudes by $A_ne^{iB_n}\equiv\alpha_n$):
\begin{align}
\frac{2\omega_{l}}{\epsilon^{2}}\frac{dA_{l}}{dt}=&-\underbrace{\sum_{i}^{\{i,j\}}\sum_{j}^{\neq}\sum_{k}^{\{k,l\}}}_{\omega_{i}+\omega_{j}=\omega_{k}+\omega_{l}}S_{ijkl}A_{i}A_{j}A_{k}\sin(B_{l}+B_{k}-B_{i}-B_{j}), \label{Eqn:RGA}\\
\frac{2\omega_{l}A_{l}}{\epsilon^{2}}\frac{dB_{l}}{dt}=&-T_{l}A_{l}^{3}-\sum_{i}^{i\neq l}R_{il}A_{i}^{2}A_{l}-\underbrace{\sum_{i}^{\{i,j\}}\sum_{j}^{\neq}\sum_{k}^{\{k,l\}}}_{\omega_{i}+\omega_{j}=\omega_{k}+\omega_{l}}S_{ijkl}A_{i}A_{j}A_{k}\cos(B_{l}+B_{k}-B_{i}-B_{j}).\label{Eqn:RGB}
\end{align}
The coefficients of the terms appearing in these time-averaged equations are expressed in terms of certain integrals of products of the AdS mode functions (\ref{scmode}):
\begin{equation}
T_{l}=\frac{1}{2}\omega_{l}^{2}X_{llll}+\frac{3}{2}Y_{llll}+2\omega_{l}^{4}W^{(0,0)}_{llll}+2\omega_{l}^{2}W^{(1,0)}_{llll},
\end{equation}
\begin{align}
R_{il}=&
\frac{1}{2}\left(\frac{\omega_{i}^{2}+\omega_{l}^{2}}{\omega_{l}^{2}-\omega_{i}^{2}}\right)\left(\omega_{l}^{2}X_{illi}-\omega_{i}^{2}X_{liil}\right)+2\left(\frac{\omega_{l}^{2}Y_{ilil}-\omega_{i}^{2}Y_{lili}}{\omega_{l}^{2}-\omega_{i}^{2}}\right)+\frac{1}{2}(Y_{iill}+Y_{llii}) \nonumber \\
+&\left(\frac{\omega_{i}^{2}\omega_{l}^{2}}{\omega_{l}^{2}-\omega_{i}^{2}}\right)\left(X_{illi}-X_{lili}\right)+\omega_{i}^{2}\omega_{l}^{2}(W^{(0,0)}_{llii}+W^{(0,0)}_{iill})+\omega_{i}^{2}W^{(1,0)}_{llii}+\omega_{l}^{2}W^{(1,0)}_{iill},
\end{align}
\begin{align}
S_{ijkl}=&
-\frac{1}{4}\left(\frac{1}{\omega_{i}+\omega_{j}}+\frac{1}{\omega_{i}-\omega_{k}}+\frac{1}{\omega_{j}-\omega_{k}}\right)(\omega_{i}\omega_{j}\omega_{k}X_{lijk}-\omega_{l}Y_{iljk}) \nonumber \\
&-\frac{1}{4}\left(\frac{1}{\omega_{i}+\omega_{j}}+\frac{1}{\omega_{i}-\omega_{k}}-\frac{1}{\omega_{j}-\omega_{k}}\right)(\omega_{j}\omega_{k}\omega_{l}X_{ijkl}-\omega_{i}Y_{jikl}) \nonumber \\
&-\frac{1}{4}\left(\frac{1}{\omega_{i}+\omega_{j}}-\frac{1}{\omega_{i}-\omega_{k}}+\frac{1}{\omega_{j}-\omega_{k}}\right)(\omega_{i}\omega_{k}\omega_{l}X_{jikl}-\omega_{j}Y_{ijkl}) \nonumber \\
&-\frac{1}{4}\left(\frac{1}{\omega_{i}+\omega_{j}}-\frac{1}{\omega_{i}-\omega_{k}}-\frac{1}{\omega_{j}-\omega_{k}}\right)(\omega_{i}\omega_{j}\omega_{l}X_{kijl}-\omega_{k}Y_{ikjl}). \label{Eqn:Sijkl}
\end{align}
The integrals in the expressions above are defined by
\begin{subequations}
\begin{align}
X_{ijkl}&=\int_{0}^{\pi/2}\text{d}x\,e'_{i}(x)e_{j}(x)e_{k}(x)e_{l}(x)(\mu(x))^{2}\nu(x), \\
Y_{ijkl}&=\int_{0}^{\pi/2}\text{d}x\,e'_{i}(x)e_{j}(x)e'_{k}(x)e'_{l}(x)(\mu(x))^{2}\nu(x), \\
W^{(0,0)}_{ijkl}&=\int_{0}^{\pi/2}\text{d}x\,e_{i}(x)e_{j}(x)\mu(x)\nu(x)\int_{0}^{x}\text{d}y\,e_{k}(y)e_{l}(y)\mu(y), \label{W00}\\
W^{(1,0)}_{ijkl}&=\int_{0}^{\pi/2}\text{d}x\,e'_{i}(x)e'_{j}(x)\mu(x)\nu(x)\int_{0}^{x}\text{d}y\,e_{k}(y)e_{l}(y)\mu(y).\label{W10}
\end{align}
\end{subequations}

The construction of quasiperiodic solutions \cite{FPU} of (\ref{Eqn:RGA}-\ref{Eqn:RGB}) is implemented analogously to the procedure we have described for (\ref{qpscalar}). One first assumes $B_n=(\beta_0+n(\beta_1-\beta_0))\eps^2 t$, whereupon (\ref{Eqn:RGA}) ensures that $A_n$ are constant (this is how quasiperiodic solutions of the sort treated here are defined). Substituting the ansatz for $B_n$ into (\ref{Eqn:RGB}), on the other hand, leads to an infinite system of algebraic equations for $\beta_0$, $\beta_1$, $A_n$, analogous to (\ref{qpscalar}).

One can furthermore concentrate on solutions strongly localized around one chosen mode, as we did in section \ref{scalarqp}, with the localization controlled by an adjustible parameter $\delta$. The resulting structure is straightforwardly analogous to (\ref{q_scalar}), with the small $\delta$ limit truncating the infinite sums in quasiperiodicity condition to finite sums, and solution for $\beta_0$, $\beta_1$ and (iteratively) $A_n$ becoming possible. In particular, the same explanation as in section \ref{scalarqp} for the existence of two spectral branches holds. (The branches, present for solutions dominated by modes other than mode 0, were originally observed numerically in \cite{FPU}.)

\subsection{Asymptotics of the interaction coefficients}

As for the probe scalar case, the asymptotic analysis of the quasiperiodicity condition for gravity involves understanding the properties of the interaction coefficients appearing in (\ref{Eqn:RGA}-\ref{Eqn:RGB}). Of particular importance are the $S$ coefficients with two or three large indices (since these are the coefficients controlling the complicated sums coupling a large number of different triplets of modes).

For $d=3$ one can apply a similar sort of analysis to what we described in section \ref{Canalysis} to extract the values of the $X$ and $Y$ integrals when three indices becomes large. For $k>0$ and $0\leqslant j\leqslant k$, one gets
\begin{align}
X_{0kj(k-j)}&=
\begin{cases}
-\frac{9\sqrt{2}}{8\pi} & \text{ if }\,\, 0<j<k \\
-\frac{9\sqrt{2}}{4\pi} & \text{ if }\,\, j=0,k
\end{cases}
\end{align}\vspace{-3mm}
\begin{align}
X_{kj(k-j)0}&=
\begin{cases}
-\frac{(75+10k)\sqrt{2}}{8\pi} & \text{ if }\,\, 0<j<k \\
-\frac{7\sqrt{2}}{\pi} & \text{ if }\,\, j=0,k
\end{cases}
\end{align}\vspace{-3mm}
\begin{align}
X_{j(k-j)k0}&=
\begin{cases}
-\frac{(45-10j)\sqrt{2}}{8\pi} & \text{ if }\,\, 0<j<k \\
-\frac{13\sqrt{2}}{4\pi} & \text{ if }\,\, j=0 \\
-\frac{7\sqrt{2}}{\pi} & \text{ if }\,\, j=k
\end{cases}
\end{align}\vspace{-3mm}
\begin{align}
Y_{k0j(k-j)}&=\frac{32\sqrt{2}}{\pi^{2}}\times
\begin{cases}
-\frac{661\pi}{256}-\frac{135\pi}{64}k+\frac{75\pi}{64}j(k-j)-\frac{45\pi}{64}k^{2}+\frac{5\pi}{32}jk(k-j) & \text{ if }\,\, 0<j<k \\
-\frac{245\pi}{128}-\frac{39\pi}{32}k-\frac{13\pi}{32}k^{2}& \text{ if }\,\, j=0,k
\end{cases}
\end{align}\vspace{-5mm}
\begin{align}
Y_{0kj(k-j)}&=-\frac{96\sqrt{2}}{\pi^{2}}\times
\begin{cases}
\frac{13\pi}{256}-\frac{13\pi}{128}k-\frac{3\pi}{64}j(k-j) & \text{ if }\,\, 0<j<k \\
\frac{5\pi}{32}& \text{ if }\,\, j=0,k
\end{cases}
\end{align}\vspace{-3mm}
\begin{align}
Y_{(k-j)jk0}&=-\frac{96\sqrt{2}}{\pi^{2}}\times
\begin{cases}
\frac{91\pi}{256}-\frac{5\pi}{128}j+\frac{9\pi}{64}k+\frac{3\pi}{64}k(k-j) & \text{ if }\,\, 0<j<k \\
\frac{59\pi}{128}+\frac{9\pi}{32}k+\frac{3\pi}{32}k^{2} & \text{ if }\,\, j=0 \\
\frac{5\pi}{32} & \text{ if }\,\, j=k
\end{cases}
\end{align}
With these values, one can compute the corresponding $S$ coefficients. One discovers, however, that the dominant contribution, which would scale like the square of the index numbers according to (\ref{Eqn:Sijkl}), in fact exactly vanishes. This leaves $S$ growing as the first power of the index number (under uniform scaling of the indices) when 3 indices become large, and when the resonance condition is satisfied.

One can perform similar analysis for $S$ with only two indices becoming large. In this case, cancellations do not occur and one discovers that $S$ grows like the index number squared (under uniform scaling of the two large indices).

The above considerations leave one with a topography of the plot of the $S$ coefficients (see Fig.~\ref{Splot}) much more complicated than the simple plateau we have observed asymptotically for the case of the probe scalar field. For example, for $S_{0kj,k-j}$, if both indices $k$ and $j$ are scaled up uniformly in proportion to $\lambda$, the $S$-coefficient also scales in proportion to $\lambda$. But if $j$ (or $k-j$) is allowed to stray close to small values, the plot rises dramatically, attaining values proportional to $k^2$.
\begin{figure}[t]
\centering
\includegraphics[width=7cm, bb=0 0 520 500]{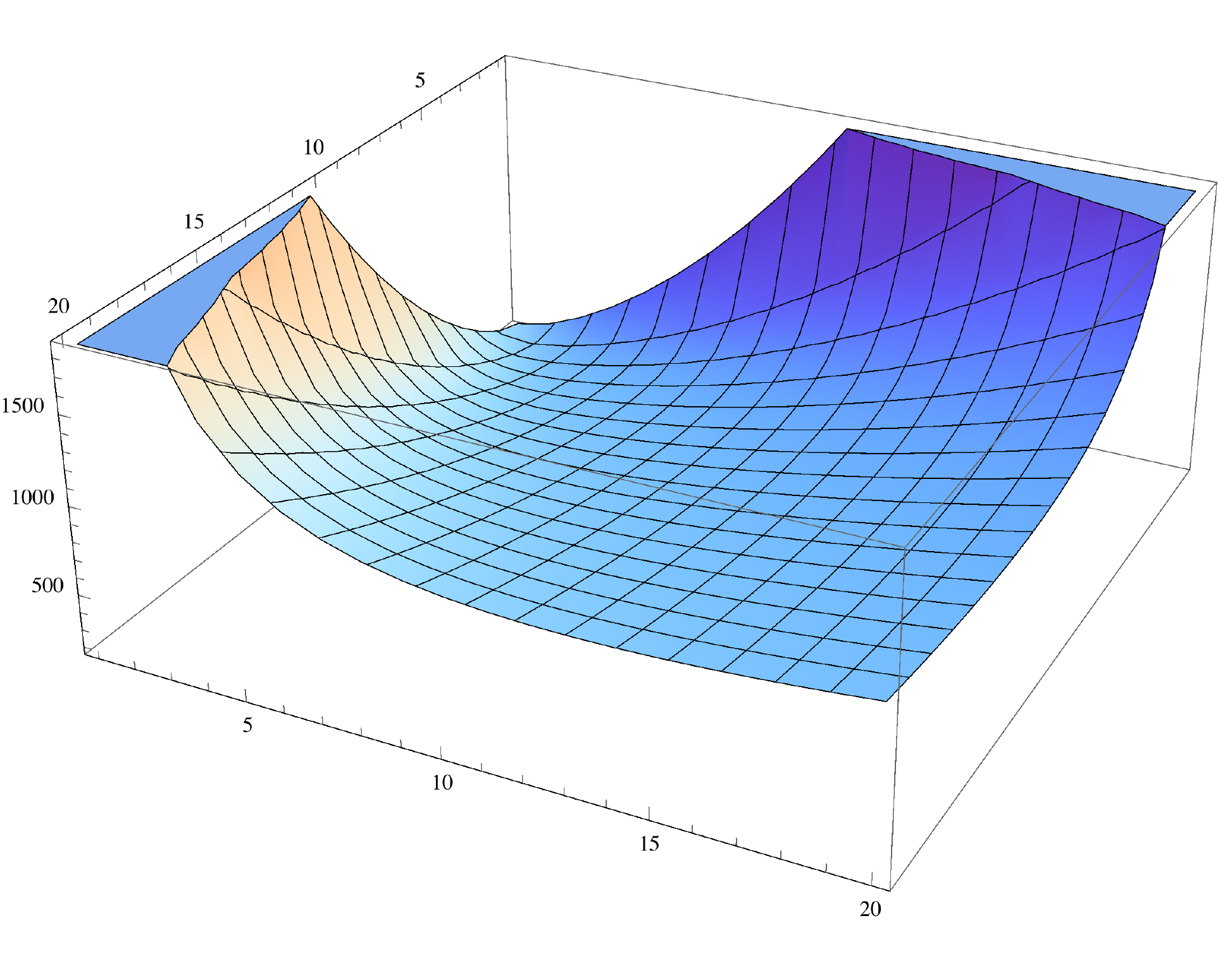}\hspace{1cm}
\includegraphics[width=7cm, bb=0 -25 520 500]{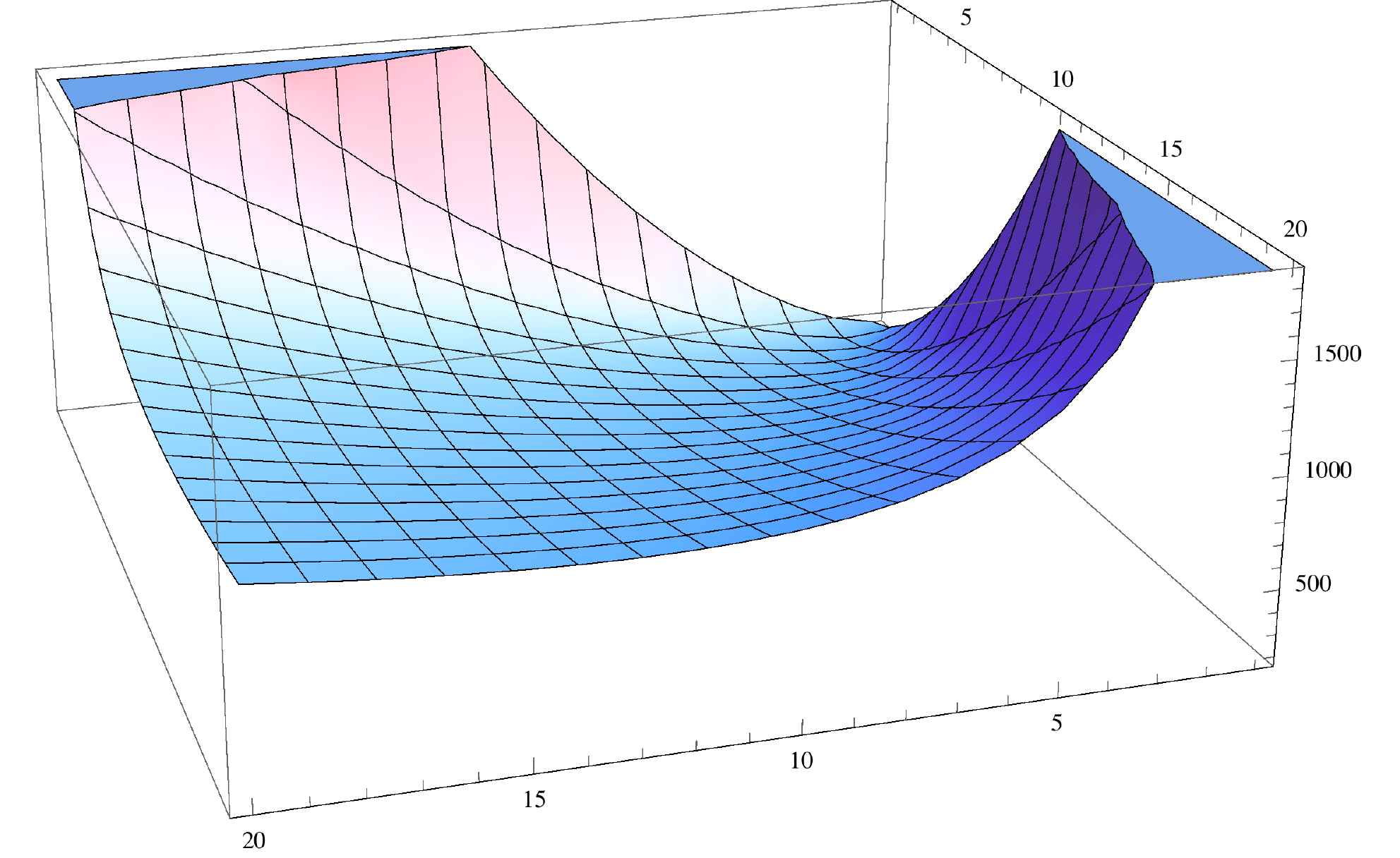}\vspace{-3mm}
\caption{$S_{0,n+m,n,m}$ plotted as a function of $n$ and $m$, viewed from two different angles.}
\label{Splot}
\end{figure}

\subsection{Solutions dominated by mode 0}

We can now describe the analytic structure of gravitational quasiperiodic solutions strongly dominated by mode number 0. Just as we did for the probe scalar field, one assumes a spectrum of the form $A_0=1$, $A_1=\de\ll 1$ and $A_n=q_n\de^n$ and substitutes it in (\ref{Eqn:RGB}), together with $B_n=(\beta_0+n(\beta_1-\beta_0))\eps^2 t$, keeping only the lowest powers of $\de$ in each equation. Equations number 0 and 1 then determine $\beta_0$ and $\beta_1$. The remaining equations take the form
\beq
\left\{2\om_k(\beta_0+k(\beta_1-\beta_0))+R_{0k}\right\}q_k=-\sum\limits_{j=1}^{k-1}S_{k-j,j0k}q_jq_{k-j}.
\label{q_grav}
\eeq
One cannot, however, proceed with this equation in the same way we did for the probe scalar field case. The reason is that the dependence of $S$ on its indices favors the summation index regions close to the end-points in (\ref{q_grav}), where $q$ itself cannot be approximated by its asymptotics. At this stage, we cannot see any obvious ways to proceed with our asymptotic analysis and extract the large mode number behavior of $q$, even if one has detailed control over the asymptotics of $S$ and $R$ (for example, via the exact analytic expressions of \cite{GMLL}).

One might be surprised that the exact asymptotics for the gravitational case proves so elusive, given that simple fitting formulas, $q_n\sim e^{\kappa n}/n$ in our language, have been proposed for solutions dominated by mode 0 in \cite{FPU}. (Note that, just like for the scalar case, $e^{\kappa n}$ drops out if substituted in (\ref{q_grav}) and it simply corrects $\delta^n$ to the actual asymptotic exponential fall-off, given that the normalization of $\de$ is fixed in the infrared by imposing $\de=A_1$.) If the asymptotic solution is as simple as $q_n\sim e^{\kappa n}/n$, it should seem strange that one cannot at least verify it using the equations and encounters all the analytic difficulties we have observed. Given this perception, we have decided to re-examine the issue numerically. Even in the absence of successful asymptotic analysis, (\ref{q_grav}) provides a straightforward iterative way to evaluate $q_n$, which can be easily programmed in Mathematica. We have implemented this procedure up to mode number 70 in hope of getting a better sense of the asymptotic behavior of $q_n$.

We first attempted to fit our numerical findings to $q_n=Qn^\gamma e^{\kappa n}$, which is what we have used for the scalar case. All of our fits are performed by choosing a group of adjacent points with the number of points equal to the number of unknowns in the putative asymptotic formula, fitting exactly to this group of points, and then moving the group of points to higher and higher index values in search of convergence. For $q_n=Qn^\gamma e^{\kappa n}$, we observed very slow convergence for $\gamma$. While not being very far from $-1$, $\gamma$ fails to reach a plateau all the way until mode number 70.

We have then attempted the same procedure with a logarithmic modulation $q_n=Qn^\gamma(\ln n)^{\gamma_1} e^{\kappa n}$ and observed that the convergence of $\gamma$ is considerably improved, while $\gamma_1$ does not appear to go anywhere close to 0. (Note that a similar procedure applied to the probe scalar field spectrum yields $\gamma_1$ rapidly converging to 0.) We finally attempted to test the presence of $\ln(\ln n)$ corrections, though admittedly our mode range is not particularly large to allow tracing very slow modulations reliably. Nonetheless, applying the fitting procedure to
\beq
q_n=Qn^\gamma(\ln n)^{\gamma_1} (\ln\ln n)^{\gamma_2}e^{\kappa n}
\left(1+\frac{c_1}{n}+\frac{c'_1}{n\ln n}+\frac{c_2}{n^2}+\frac{c'_2}{n^2\ln n}+\frac{c''_2}{n^2\ln^2 n}\right)
\eeq
results in good convergence within the examined mode range. $\gamma$ rapidly reaches $-1$, within a percent, while $\gamma_1$ and $\gamma_2$ appear to be around $0.3$ and $0.1$, respectively. (These values change by only a few percent between estimating around mode 50 and around mode 70 with our procedure, indicating adequate convergence. Removing the corrections decreasing as $1/n^2$ and faster from the above ansatz makes the plots less flat, though the actual estimated values for the powers remain close to the ones quoted above.) We thus find evidence that quasiperiodic spectra of AdS perturbations with gravitational non-linearities display slow logarithmic modulations in their ultraviolet asymptotics, making them considerably more challenging for both analytic and numerical treatments than the power laws previously mentioned in the literature. It would be interesting to analyze such logarithmic corrections further, especially with the explicit expressions for the interaction coefficients in AdS$_4$ that have just appeared in \cite{GMLL}. The numerical optimization allowed by these expressions makes it possible to go to much higher mode numbers than what one can get by brute force evaluation. (See also our parallel work \cite{CEV3} that presents effective iterative techniques for evaluating the interaction coefficients in a general number of dimensions.)


\section{Acknowledgments}

We would like to thank Piotr Bizo\'n, Chethan Krishnan, Luis Lehner, Maciej Maliborski and Andrzej Rostworowski for useful discussions.
The work of B.C.\ and J.V.\ has been supported by the Belgian Federal Science Policy Office through the Interuniversity Attraction Pole P7/37, by FWO-Vlaanderen through project G020714N, and by the Vrije Universiteit Brussel through the Strategic Research Program ``High-Energy Physics.'' The work of O.E. is funded under CUniverse
research promotion project by Chulalongkorn University (grant reference CUAASC). P.J. is supported by a scholarship under the Development and Promotion of Science and
Technology Talents Project (DPST) by the government of Thailand.
J.V.\ is supported by a PhD fellowship of Research Foundation Flanders (FWO).



\end{document}